\documentclass[%
 aip,
 amsmath,amssymb,
 reprint,%
]{revtex4-1}
\usepackage{graphicx}
\usepackage{dcolumn}
\usepackage{bm}

\usepackage[utf8]{inputenc}
\usepackage[T1]{fontenc}
\usepackage{mathptmx}
\usepackage{etoolbox}

\usepackage{algpseudocode}     
\usepackage{amsmath} 
\usepackage{array}
\usepackage{xcolor}
\usepackage{diagbox}
\makeatletter
\def\@email#1#2{%
 \endgroup
 \patchcmd{\titleblock@produce}
  {\frontmatter@RRAPformat}
  {\frontmatter@RRAPformat{\produce@RRAP{*#1\href{mailto:#2}{#2}}}\frontmatter@RRAPformat}
  {}{}
}%
\makeatother
\begin{document}

\preprint{AIP/123-QED}

\title[Rapid and robust laser-frequency auto-locking...]{Rapid and robust laser-frequency auto-locking using Bayesian-optimization and discrete-wavelet-transformation algorithms}

\author{Min Jiang}
\affiliation{State Key Laboratory of Magnetic Resonance and Atomic and Molecular Physics, Innovation Academy for Precision Measurement Science and Technology, Chinese Academy of Sciences, Wuhan 430071, China}

\author{Xiao-Li Chen}
\affiliation{State Key Laboratory of Magnetic Resonance and Atomic and Molecular Physics, Innovation Academy for Precision Measurement Science and Technology, Chinese Academy of Sciences, Wuhan 430071, China}

\author{Si-Bin Lu\textsuperscript{*}}
\affiliation{State Key Laboratory of Magnetic Resonance and Atomic and Molecular Physics, Innovation Academy for Precision Measurement Science and Technology, Chinese Academy of Sciences, Wuhan 430071, China}

\author{Jia-Hao Fu}
\affiliation{State Key Laboratory of Magnetic Resonance and Atomic and Molecular Physics, Innovation Academy for Precision Measurement Science and Technology, Chinese Academy of Sciences, Wuhan 430071, China}
\affiliation{School of Physics, University of Chinese Academy of Sciences, Beijing 100049, China}

\author{Zhan-Wei Yao\textsuperscript{*}}
\affiliation{State Key Laboratory of Magnetic Resonance and Atomic and Molecular Physics, Innovation Academy for Precision Measurement Science and Technology, Chinese Academy of Sciences, Wuhan 430071, China}
\affiliation{Hefei National Laboratory, Hefei 230088, China} 

\author{Shao-Kang Li}
\affiliation{State Key Laboratory of Magnetic Resonance and Atomic and Molecular Physics, Innovation Academy for Precision Measurement Science and Technology, Chinese Academy of Sciences, Wuhan 430071, China}

\author{Min Ke}
\affiliation{State Key Laboratory of Magnetic Resonance and Atomic and Molecular Physics, Innovation Academy for Precision Measurement Science and Technology, Chinese Academy of Sciences, Wuhan 430071, China}

\author{San-Ming Song}
\affiliation{Shenzhen Institute of Advanced Technology, Chinese Academy of Sciences, Shenzhen 518055, China}

\author{Run-Bing Li\textsuperscript{*}}
\affiliation{State Key Laboratory of Magnetic Resonance and Atomic and Molecular Physics, Innovation Academy for Precision Measurement Science and Technology, Chinese Academy of Sciences, Wuhan 430071, China}
\affiliation{Hefei National Laboratory, Hefei 230088, China}
\affiliation{Wuhan Institute of Quantum Technology, Wuhan 430206, China}

\author{Jin Wang}
\affiliation{State Key Laboratory of Magnetic Resonance and Atomic and Molecular Physics, Innovation Academy for Precision Measurement Science and Technology, Chinese Academy of Sciences, Wuhan 430071, China}
\affiliation{Hefei National Laboratory, Hefei 230088, China}
\affiliation{Wuhan Institute of Quantum Technology, Wuhan 430206, China}

\author{Ming-Sheng Zhan}
\affiliation{State Key Laboratory of Magnetic Resonance and Atomic and Molecular Physics, Innovation Academy for Precision Measurement Science and Technology, Chinese Academy of Sciences, Wuhan 430071, China}
\affiliation{Hefei National Laboratory, Hefei 230088, China}
\affiliation{Wuhan Institute of Quantum Technology, Wuhan 430206, China}

\email[]{Authors to whom correspondence should be addressed: lusibin@apm.ac.cn, yaozhw@apm.ac.cn, rbli@wipm.ac.cn}

\date{\today}
\begin{abstract}
Rapid and robust laser-frequency auto-locking is essential for the field deployment of quantum communications, quantum computing, and precision-measurement technologies; however, achieving this remains a considerable challenge. Here, we propose and demonstrate an auto-locking scheme employing Bayesian optimization and discrete biorthogonal wavelet transformation. First, the reference is rapidly sought by making intelligent use of historical observations, eliminating the inherent blindness of the traditional parameter-scanning method. Second, the frequency reference is robustly identified by pinpointing transition signals with the discrete biorthogonal wavelet transformation and analyzing their immutable frequency differences and relative magnitudes, which are determined by the inherent atomic structure and remain resistant to environmental disturbances. This proposed approach achieves a fivefold acceleration in reference searching compared to conventional scanning methods in the case where the laser frequency drifts far away from the reference. Crucially, it achieves an identification accuracy of more than 99.5\%, even under severe 50\% laser-intensity fluctuations, $9.95^\circ$ photodiode misalignment, and $18^\circ$C Rb cell temperature elevation. Finally, locking the laser frequency to the identified reference with a lead zirconate titanate-current double-servo loop narrows the linewidth to 20~kHz. We believe that this rapid, robust, and high-performance auto-locking technique will be pivotal towards the deployment of the next generation of practical quantum technologies in demanding field environments.
\end{abstract}

\maketitle

\section{\label{sec:introduction}INTRODUCTION}
Laser-frequency locking plays a critical role in a variety of quantum technologies, including quantum computing~\cite{chen2022,pogorelov2021}, quantum communication~\cite{miyashita2021,bulla2023nonlocal}, and quantum precision-measurement systems~\cite{Kasevich2011Absolute,yao2021self,fang2018realization,zhu2022feedback}. As these technologies progressively transition from the laboratory to field applications~\cite{schleich2016quantum,Bongs2019,Narducci2022}---for example through deployment in space stations~\cite{Elliott2023,He2023} and sounding rockets~\cite{dinkelaker2017}---laser-frequency auto-locking will become increasingly important. Although various auto-locking methods have been developed~\cite{allard2004,koch2003,ruksasakchai2022,subhankar2019,luo2016,zhang2020,yan2022,guo2022,berthiaume2010,dinkelaker2017,li2020}, their practical use is constrained by slow reference search speeds and limited robustness in reference identification. To enable broader field deployment, there is an urgent need for laser-frequency auto-locking methods that can quickly locate and reliably lock on to a reference, thereby ensuring the immediate readiness of quantum systems and improving their adaptability to changing environmental conditions.

\begin{figure*}[htb]
\includegraphics[scale=0.5]{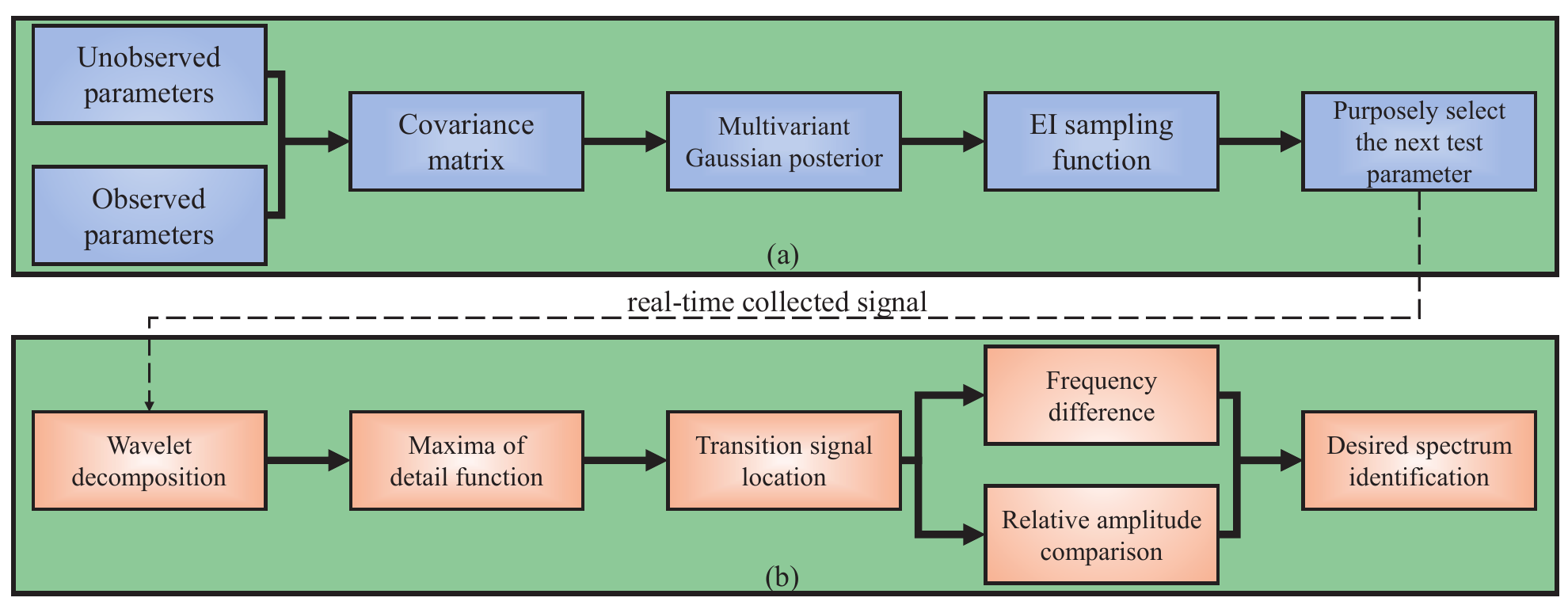}
\caption{\label{fig:Principle}(Color online) Framework of the proposed scheme. (a)~The key is to use the multivariate Gaussian posterior with expected-improvement (EI) sampling to deliberately select the next test parameter, and to use the maxima of the detail function, (b)~derived from the wavelet decomposition, to locate the transition signal in the real-time collected signal.}
\end{figure*}

First, rapidly searching for the reference is an essential capability for field-deployable quantum technologies, particularly in applications such as quantum inertial navigation, where continuous sensor operation is preferable. Compared with conventional gradual-scanning methods that use a fixed scanning interval and fixed step size, the mean-shift algorithm, which can dynamically calculate the initial scanning value~\cite{fan2022}, and the coarse-to-fine search strategy, which applies a large step size followed by a refined small step size~\cite{yan2022}, can improve search efficiency. Despite these advances, such approaches have not fundamentally changed the gradual-scanning nature of the process, and this leads to inherent limitations. In particular, when the search range is large, the search efficiency is still low.

Second, in addition to searching of the reference, accurately and robustly identifying that reference constitutes another indispensable requirement for quantum systems employed in field applications. Threshold comparison methods~\cite{allard2004,koch2003,ruksasakchai2022,subhankar2019,luo2016,zhang2020,yan2022,guo2022} and cross-correlation calculation algorithms~\cite{berthiaume2010,dinkelaker2017,li2020} can be used to accurately identify the reference within specific contexts, but they may misidentify the reference when the laser intensity is affected by environmental variations such as temperature fluctuations~\cite{durfee2006}. Gaussian continuous-wavelet-transform-based noise reduction~\cite{li2024} can improve robustness, but it does so at the expense of altering the signal. In a similar trade-off, a deep-neural-network identifier~\cite{winkler2023} has been shown to provide robust reference identification, but it requires 80~s of processing time. To our best knowledge, no existing method can both rapidly search for and robustly identify the reference signal simultaneously.

In this paper, we present a rapid and highly robust laser-frequency auto-locking scheme using Bayesian-optimization and discrete-wavelet-transformation algorithms. First, a Gaussian-process-based Bayesian-optimization method is employed to speed up the reference search. By purposefully selecting the next test value based on historical observations of the laser-frequency control parameters, this approach avoids the blindness inherent in conventional gradual-parameter-scanning methods. A discrete biorthogonal wavelet transformation is then used to accurately locate atomic transition signals within the collected spectra. By assessing the frequency differences and relative magnitudes of these atomic transition signals, the robustness of reference identification is strengthened. Our experimental results show that the Gaussian-process-based Bayesian-optimization algorithm achieves a fivefold increase in reference searching speed compared to the conventional gradual-scanning method, especially when the laser frequency drifts far away from the reference, and the discrete biorthogonal wavelet-transformation method achieves a 99.5\% reference-identification rate, even under 50\% laser-intensity variation, $9.95^\circ$ photodiode misalignment, and $18^\circ$C Rb cell temperature elevation. Finally, the laser frequency is auto-locked by rapidly searching for and accurately identifying the atomic spectrum, reducing the laser linewidth to 20~kHz using a combined current and lead-zirconate-titanate (PZT) double-servo feedback control loop. This rapid and robust laser-frequency auto-locking method with narrow-linewidth performance is thus beneficial for advancing quantum technologies towards field applications.

\section{\label{sec:Methodology}PRINCIPLE AND METHOD}
Figure~\ref{fig:Principle} shows the framework of an advanced scheme for rapidly searching for and accurately identifying reference signals. The reference signal is rapidly sought using Gaussian-process-based Bayesian optimization. As illustrated in Fig.~\ref{fig:Principle}(a), the observed laser-frequency control parameters are used to calculate the covariance with unobserved parameters. The predicted signal values and uncertainties of the unobserved parameters are provided by the multivariate Gaussian posterior. With this posterior, an expected-improvement (EI) sampling function is used to estimate the next parameter at which the maximum or minimum signal value, corresponding to the atomic transition, may occur. Let $x$ denote the laser-frequency control parameter and its set be $\boldsymbol{\chi}$. The functional relationship between the reference signal and the parameter $x$ is $g(x)$, which is unknown because the laser frequency may drift due to variations in environmental conditions. Gaussian processes offer a good balance between modeling accuracy and computational complexity; thus, the Gaussian process $f(x)$ is used as an alternative model for $g(x)$. Let $n$ observed parameter values $x_{i}\in \boldsymbol{\chi}$ form a vector $\bm{X} = \{x_i\}_{i = 1}^n$. The corresponding observed signal values ${f_i}{\buildrel \Delta \over =}f(x_i)$ form the vector $\bm{f} = \{f_i\}_{i = 1}^n$. The covariance matrix between the two parameter vectors $\bm{X}^1$ and $\bm{X}^2$ is defined as
\begin{equation}\label{eq:1}
\bm{K}\left( {\bm{X^1},\bm{X^2}} \right) = \left[ {\begin{array}{*{20}{c}}
{k\left( {x_1^1,x_1^2} \right)}&{k\left( {x_1^1,x_2^2} \right)}& \cdots &{k\left( {x_1^1,x_n^2} \right)}\\
 \vdots & \vdots & \ddots & \vdots \\
{k\left( {x_m^1,x_1^2} \right)}&{k\left( {x_m^1,x_2^2} \right)}& \ldots &{k\left( {x_m^1,x_n^2} \right)}
\end{array}} \right],
\end{equation}
where $m$ and $n$ are the numbers of elements in the vectors $\bm{X^1}$ and $\bm{X^2}$, and $k\left(x_p,x_q\right)\allowbreak=\exp\left(-\left|x_p-x_q\right|^2/2\right)$. Let $\bm{X_*}$ denote a vector of unobserved parameters and $\bm{f_*}$ the corresponding unobserved function values. Since $g(x)$ is modeled by the Gaussian process $f(x)$, the random variables $\bm{f}$ and $\bm{f_*}$ follow a multivariate Gaussian distribution~\cite{bishop2006,maggi2021}:
\begin{equation}\label{eq:2}
\left[ {\begin{array}{*{20}{c}}
\bm{f}\\
\bm{f_*}
\end{array}} \right] \; = {\cal N}\left( {\begin{array}{*{20}{c}}
\bm{0}, & {\left[ {\begin{array}{*{20}{c}}
\bm{K}(\bm{X},\bm{X}) & \bm{K}(\bm{X},\bm{X_*})\\
\bm{K}(\bm{X_*},\bm{X}) & \bm{K}(\bm{X_*},\bm{X_*})
\end{array}} \right]}
\end{array}} \right).
\end{equation}
According to the conditional probability formula of multivariate Gaussian functions, the posterior probability density function of the Gaussian process is given by
\begin{equation}\label{eq:3}
\begin{split}
\bm{f_*} &\mid \bm{X_*}, \bm{X}, \bm{f} \sim \mathcal{N}\Big(
\bm{K}(\bm{X_*},\bm{X}) \, \bm{K}(\bm{X},\bm{X})^{-1} \, \bm{f}, \\
&\bm{K}(\bm{X_*},\bm{X_*}) - \bm{K}(\bm{X_*},\bm{X}) \, \bm{K}(\bm{X},\bm{X})^{-1} \, \bm{K}(\bm{X},\bm{X_*})
\Big)
\end{split}
\end{equation}

To find the maximum or minimum signal value, the next test value $x_{n+1}$ is purposefully selected by the EI sampling function based on the mean and variance of $\bm{f_*}|\bm{X_*},\bm{X},\bm{f}$ and the currently observed maximum or minimum value of $f(x)$. The following formula shows the maximum signal value search; the minimization case is analogous:
\begin{equation}\label{eq:4}
x_{n + 1} = \mathop{\text{arg}\max}\limits_{x \in \chi} E{\left[ {\left( f(x) - \mathop{\max}\limits_{i = 1,\ldots,n} f(x_i) \right) \left| \bm{f} \right.} \right]^+ },
\end{equation}
where $[\cdot]^+ = \max(\cdot,0)$. Based on Eqs.~(\ref{eq:1})$\--$(\ref{eq:4}), the next laser-frequency control parameter can be purposefully selected using observed parameters and their corresponding signal values, rather than through traditional step-by-step scanning. This significantly improves the efficiency of the search, as will be experimentally validated in Section~\ref{sec:Experiments}.

Accurately identifying a desired reference signal, such as an atomic transition frequency or the characteristic frequency of an optical cavity, is indispensable in laser-frequency auto-locking. As shown in Fig.~\ref{fig:Principle}(b), in our approach, the reference signal is accurately identified using discrete wavelet transformations. Because the frequency differences and relative amplitudes of the atomic transition signals within the spectrum are determined by the atomic energy-level structure and transition probabilities, the reference signal can be robustly identified, even in the presence of environmental variations. When the laser frequency coincides with an atomic transition frequency, the collected signal will change abruptly, indicating a specific atomic transition. The discrete biorthogonal wavelet transformation can accurately locate these abrupt changes without artificial shifting, owing to the perfect symmetry of its filter coefficients. Mathematically, let $\widetilde{\phi}_{j,k}(t) = 2^{j/2}\widetilde{\phi}(2^j t - k)$ denote the biorthogonal scaling function at scale $j$ and position $k$. Here, a B-spline function of degree $n$, denoted as $B_n(t)$, is chosen as the fundamental scaling function $\widetilde{\phi}(t)$~\cite{ruch2009}. The real-time collected signal $f(t)$ can then be expressed as an expansion in this scaling space:
\begin{equation}
f(t) = \sum_{k \in \mathbb{Z}} \widetilde{a}_{j,k} \widetilde{\phi}_{j,k}(t),
\end{equation}
where $\widetilde{a}_{j,k}$ are the scaling coefficients, and $\mathbb{Z}$ is the set of integers. For a chosen degree $n$, the discrete spline scaling filter $\widetilde{h}$ can be derived analytically using the following formula:
\begin{equation}
\widetilde{h}_k = \frac{\sqrt{2}}{2^{n+1}}
\begin{cases} 
\binom{n+1}{\frac{n+1}{2} - k}, & k = -\frac{n+1}{2}, \dots, \frac{n+1}{2} \quad \text{for odd } n \\[2ex]
\binom{n+1}{\frac{n}{2} + k}, & k = -\frac{n}{2}, \dots, \frac{n}{2} + 1 \quad \text{for even } n
\end{cases}
\end{equation} 
with all other coefficients $\widetilde{h}_k = 0$, where $\binom{n}{k}$ denotes the standard binomial coefficient. The symmetric dual scaling filter $h$, which pairs with the spline filter $\widetilde{h}$, can be subsequently constructed to satisfy the biorthogonality symbol condition:
\begin{equation}
\widetilde{H}(\omega)\overline{H(\omega)} + \widetilde{H}(\omega + \pi)\overline{H(\omega + \pi)} = 1
\end{equation} 
for all $\omega \in \mathbb{R}$, where $H(\omega) = \frac{1}{\sqrt{2}}\sum_{k \in \mathbb{Z}} h_k e^{-ik\omega}$, $\widetilde{H}(\omega) = \frac{1}{\sqrt{2}}\sum_{k \in \mathbb{Z}} \widetilde{h}_k e^{-ik\omega}$, and  $\overline{H\left(\cdot\right)}$ denotes the complex conjugate of ${H\left(\cdot\right)}$. 
To accurately locate atomic transition signals within the collected spectrum, a discrete wavelet decomposition is performed on $f(t)$ using the derived dual filters $h$ and $g$: 
\begin{equation}
\label{waveletDecomposition}
\begin{split}
f(t) =& \sum_{k \in \mathbb{Z}} \left( \sum_l h_{l - 2k} \widetilde{a}_{j,l} \right) \widetilde{\phi}_{j-1,k}(t) \\ 
&+ \sum_{k \in \mathbb{Z}} \left( \sum_p g_{p - 2k} \widetilde{a}_{j,p} \right) \widetilde{\psi}_{j-1,k}(t).
\end{split}
\end{equation}
Here, $\widetilde{\psi}_{j-1,k}(t) = 2^{(j-1)/2}\widetilde{\psi}(2^{j-1}t-k)$ represents the wavelet function, defined by $\widetilde{\psi}(t) = \sqrt{2}\sum_{k \in \mathbb{Z}} \widetilde{g}_k \widetilde{\phi}(2t-k)$, and $g_k = (-1)^k\widetilde{h}_{1-k}$ serves as the dual wavelet filter.  
Iterative wavelet decomposition can be applied to the first term, i.e., the low-frequency approximation, on the right-hand side of Eq.~\ref{waveletDecomposition}, resulting in a multiscale representation:
\begin{equation}
f(t) = \sum_{m \in \mathbb{Z}} b_m \widetilde{\phi}_{r,m}(t) + \sum_{n = r}^{j - 1} \sum_{i \in \mathbb{Z}} c_{n,i} \widetilde{\psi}_{n,i}(t),
\end{equation}
where $r < j$ is the final coarse resolution level, and $b_m$ and $c_{n,i}$ represent the resulting scale and detail coefficients, respectively. When an atomic transition occurs, the values in the detail function $\sum\nolimits_{i \in \mathbb{Z}} {{c_{n,i}}{\widetilde \psi _{n,i}}(t)} $ increase. The $N$ largest values in the detail function are selected, and the density-based spatial clustering of applications with noise (DBSCAN) method~\cite{ester1996}, which does not require predefined numbers of clusters, clusters the $x$ coordinates of these points. For each cluster, zero-crossing or local-maximum points of the signal are sought. The frequency differences of the atomic transition signals are calculated based on the $x$ coordinates of these points. By also comparing the maximum and minimum values within the signal range corresponding to the clusters in the detail function, the desired spectrum and reference are robustly identified, as will be experimentally validated in the next section.  

\begin{figure}[htb]
\includegraphics[width=\columnwidth]{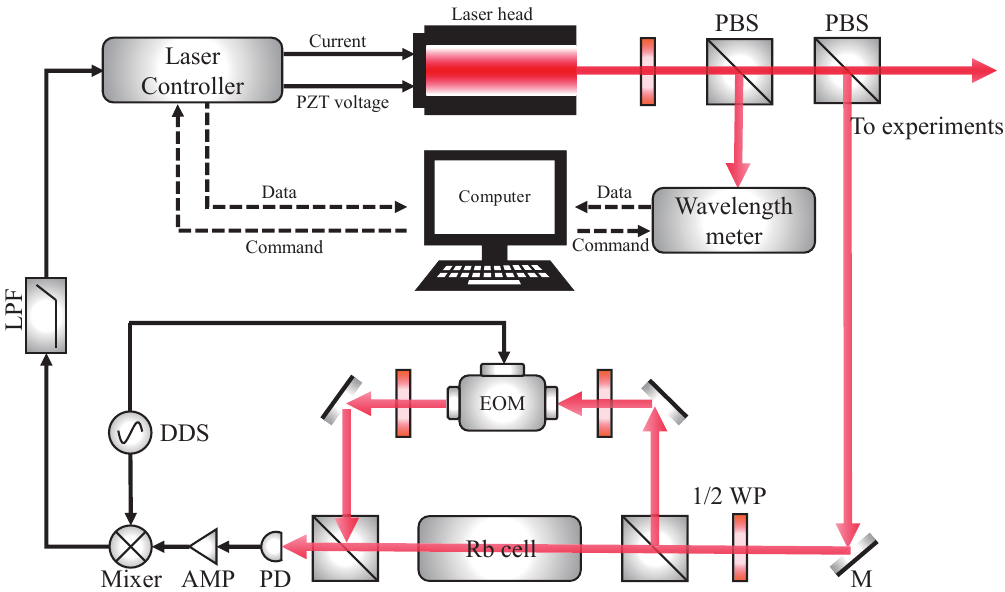}
\caption{\label{fig:Experimental setup}(Color online) Schematic of the experimental setup. The laser frequency is auto-locked using the MTS signal. Definitions: PBS, polarizing beam splitter; M, mirror; EOM, electro-optic modulator; PD, photodiode; AMP, circuit amplifier; Mixer, phase detector; DDS, direct digital frequency synthesizer; LPF, low-pass filter; 1/2 WP, half-wave plate. The laser controller includes a DigiLock~110 module, among other components, produced by TOPTICA Photonics Inc. Electrical connections are shown as black lines, with dashed black lines representing digital communication between the control computer, laser controller, and wavelength meter. The laser optical path is indicated by thick red lines.}
\end{figure}

\begin{figure}[htbp]
\hrule\vspace{3pt}
\noindent\textbf{Algorithm 1} Rapid and robust laser-frequency auto-locking
\vspace{3pt}\hrule\vspace{3pt}
\begin{algorithmic}[1]
    \Require $S(v)$, search space $\mathcal{V}$, threshold $\epsilon$, max iters $n$.
    \Ensure Laser locked to the target transition frequency.
    
    \State Initialize observation set $\mathcal{O} \gets \emptyset$;
    \State Randomly sample initial parameter $v_{next} \in \mathcal{V}$;
    
    \For{$i = 1$ \textbf{to} $n$}
        \State Set parameter to $v_{next}$ and acquire signal $S(v_{next})$;
        
        \If{$\max(|S(v_{next})|) > \epsilon$} \Comment{Bypass flat noise floor}
            \State \textbf{Wavelet Decomposition:}
            \State Decompose $S(v_{next})$ using dual filters $(h,g)$;
            \State Compute detail functions at target scales;
            
            \State \textbf{Feature Extraction:}
            \State Extract local maxima from absolute detail functions;
            \State Cluster maxima coordinates using DBSCAN;
            \State Locate candidate transition frequencies per cluster;
            \State Extract signal extrema (maxima/minima) per cluster;
            
            \State \textbf{Reference Identification:}
            \State Compute transition frequency differences $\Delta f$;
            \State Compute relative amplitudes $\Delta A$ of the extrema;
            \State Verify $\Delta f$ and $\Delta A$ against atomic parameters;
            
            \If{Target spectrum is positively identified}
                \State Engage PID lock at target transition frequency;
                \State \textbf{break};
            \EndIf
        \EndIf
        
        \State $\mathcal{O} \gets \mathcal{O} \cup \{ (v_{next}, \max|S(v_{next})|) \}$;
        \State Update Gaussian Process (GP) model using $\mathcal{O}$;
        \State Select next $v_{next} \in \mathcal{V}$ by maximizing the EI function;
    \EndFor 
\end{algorithmic}

\vspace{3pt}\hrule
\end{figure}

\section{\label{sec:Experiments}EXPERIMENTAL SETUP AND RESULTS}
\begin{figure*}[htb]
\includegraphics[scale=0.5]{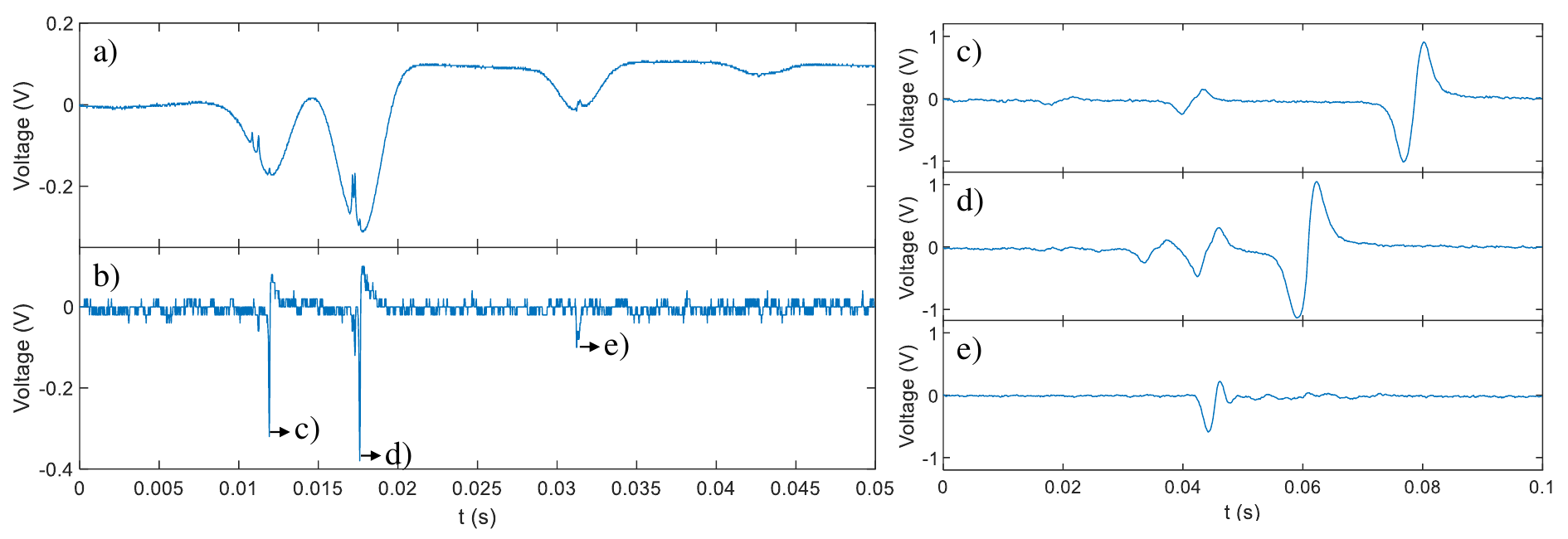}
\caption{\label{fig:intensityVariations}
(a)~Doppler-free saturated absorption spectrum and (b)~corresponding MTS signal for D2 line transitions of Rubidium atoms. (c)--(e)~Enlarged views of the transition signals marked in~(b).}
\end{figure*}

To demonstrate the effectiveness of the proposed scheme, a laser-frequency auto-locking experimental apparatus based on the modulation transfer spectrum (MTS) was set up as shown in Fig.~\ref{fig:Experimental setup}. In this apparatus, the output of the laser (an external cavity diode laser) is split into three beams after passing through two polarizing beam splitters (PBSs). One beam is directed into a wavelength meter to judge whether the laser frequency is locked to the desired frequency reference. Another beam serves as the input to the subsequent optical path, and the third beam, which is reflected by a mirror, is again split into the probe and pump beams after passing through a half-wave plate and a PBS. The polarization of the pump beam is controlled by two half-wave plates located next to the electro-optic modulator (EOM). After phase modulation by the EOM, the pump beam propagates in a direction opposite to and overlapping with the probe beam in a rubidium cell. The beat signal of the probe beam is detected by a fast photodiode. After amplification, this signal is mixed with the modulation signal of the pump beam. The output of the mixer is filtered by a low-pass filter, and the MTS error signal is finally obtained, as shown in Fig.~\ref{fig:intensityVariations}(b), with the corresponding saturated absorption spectrum (SAS) shown in Fig.~\ref{fig:intensityVariations}(a). Figures~\ref{fig:intensityVariations}(c)--\ref{fig:intensityVariations}(e) show enlarged sections of the transition signals marked in Fig.~\ref{fig:intensityVariations}(b). The MTS spectrum is composed of $F=2 \to F^\prime=3$, $F=2 \to F^\prime=\text{CO} 2,3$, and $F=2 \to F^\prime=\text{CO} 1,3$ transition signals of $^{87}$Rb atoms in Fig.~\ref{fig:intensityVariations}(c); $F=3 \to F^\prime=4$, $F=3 \to F^\prime=\text{CO} 3,4$, and $F=3 \to F^\prime=\text{CO} 2,4$ transition signals of $^{85}$Rb atoms in Fig.~\ref{fig:intensityVariations}(d); and $F=2 \to F^\prime=3$, $F=2 \to F^\prime=\text{CO} 2,3$, and $F=2 \to F^\prime=\text{CO} 1,3$ transition signals of $^{85}$Rb atoms in Fig.~\ref{fig:intensityVariations}(e). Here, $\text{CO}$ indicates crossover resonance. Note that the spectrum of transitions from the ground state of $^{87}$Rb $F=1$ to the excited states is not shown due to its small signal amplitude in the current experimental apparatus. Figure~\ref{fig:intensityVariations}(c)~is taken as the desired spectrum, and the transition frequency of $F=2 \to F^\prime=3$ is used as the frequency reference in the subsequent experiments. The reference-identification algorithms should be able to distinguish the desired spectrum from the others, especially when their shapes are very similar.

\begin{figure}[htb]
\includegraphics[width=\columnwidth]{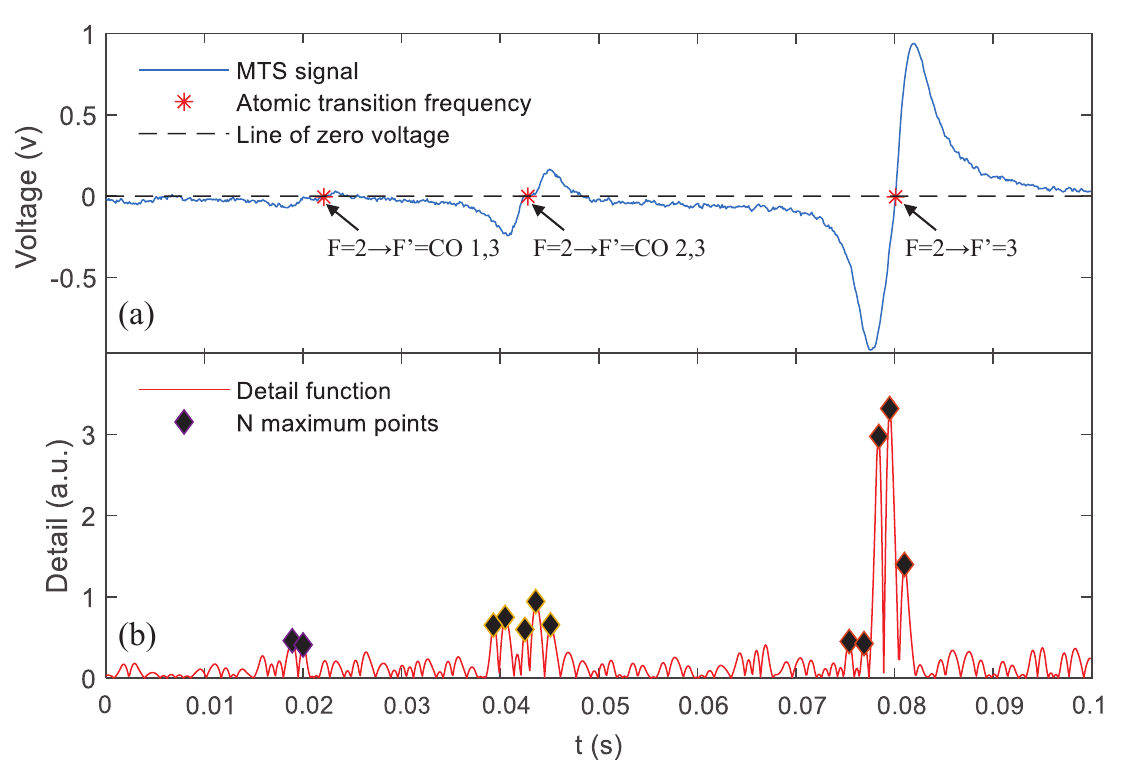}
\caption{\label{fig:wavelet}(Color online) Accurate localization of abrupt changes of the MTS signal using wavelet transforms. The asterisk on the far right in the desired spectrum indicates the reference being locked to~(a), and the $N=15$ maximum-value points marked by the black diamonds from the detail function in~(b) are used to locate the transition signals.}
\end{figure}

The appropriateness of the algorithm parameter settings determines the performance of the proposed reference-identification method, and the implementation details of the algorithms are presented in Algorithm 1. To select the best parameters, the incident laser power of the MTS apparatus was first adjusted to 17.6~mW to ensure that the MTS error signal exhibited a good signal-to-noise ratio (SNR). A total of 1000~MTS error signals, as shown in Fig.~\ref{fig:intensityVariations}(c), were collected and then used to optimize the algorithm parameters to achieve the highest desired spectrum-identification rate. Specifically, a frequency window of 310~MHz was first set to encompass the three atomic transition signals of interest. Additionally, to facilitate a fast search for the desired spectrum, spectrum identification was performed only when the maximum absolute value of the signal within the frequency window exceeded 0.2~V. Selecting the degree of B-spline function in the biorthogonal wavelet transformation~\cite{ruch2009} involves a critical mathematical trade-off: higher-degree splines provide more vanishing moments to suppress noise, but their wider compact support smears the narrow transition signal.
\begin{table}[bp]
\caption{Identification accuracies of the proposed method using different biorthogonal filter pairs under varying photodiode misalignments. Increasing the misalignment angle progressively degrades the SNR.}
\label{tab:justification}
\begin{tabular*}{\linewidth}{@{\extracolsep{\fill}}lccc}
\hline\hline
Filter pair & S1\textsuperscript{a} & S2\textsuperscript{b} & S3\textsuperscript{c} \\
\hline
Linear (5,3) & 100\% & 98.7\% & 50.1\% \\
Quadratic (8,4) & 99.9\% & 99.7\% & 53.6\% \\
Cubic (7,5) & 86.5\% & 84.0\% & 38.5\% \\
\hline\hline
\end{tabular*}

\vspace{3pt}
\raggedright
\footnotesize
\textsuperscript{a} Normal alignment condition.\\
\textsuperscript{b} PD misaligned by $9.95^\circ$.\\
\textsuperscript{c} PD misaligned by $11.54^\circ$.
\end{table}
Table \ref{tab:justification} demonstrates this trade-off using photodiode misaglignments of $0^\circ$ (S1), $9.95^\circ$ (S2), and $11.54^\circ$ (S3) as low-SNR tests (which will be further detailed later in this section). The shorter (5,3) filter lacks sufficient vanishing moments to suppress severe detector noise, causing its accuracy to degrade in the S2 and S3 conditions. Conversely, the wider (7,5) filter over-smooths the signal, drastically reducing accuracy even under the clean S1 condition. 
Because the quadratic (8,4) filter perfectly balances noise suppression and sharp feature localization to yield the highest overall accuracy, the (8,4) biorthogonal filter pair was selected to gradually decompose the collected signals six times. As shown in Fig.~\ref{fig:wavelet}(b) by black diamonds, 15~maximum-value points were selected from the detail function, which was obtained by wavelet decomposition and is indicated by the red line. The DBSCAN clustering algorithm was used to cluster the $x$ coordinates of these points. The clustering parameters, including the epsilon neighborhood and minimum number of neighbors required for a core point, were set to 19.4~MHz and 2, respectively, allowing the selected maximum-value points corresponding to the three atomic transition signals to be clustered into three categories. The average of each cluster was then calculated. Using this average as the center, the zero-crossing coordinates marked by the asterisks in Fig.~\ref{fig:wavelet}(a), as well as the maxima and minima of the MTS error signals within $\pm 19.4$~MHz, were recorded, and the frequency differences between each pair of zero-crossing points were calculated. To identify the desired spectrum, the ranges of frequency differences between the $F=2 \to F^\prime=3$ and $F=2 \to F^\prime=\text{CO} 2,3$, $F=2 \to F^\prime=\text{CO} 2,3$ and $F=2 \to F^\prime=\text{CO} 1,3$, and $F=2 \to F^\prime=3$ and $F=2 \to F^\prime=\text{CO} 1,3$ transition signals were set to 113--152, 66--105, and 206--245~MHz, respectively. The calculated frequency differences between the zero-crossing points of the MTS error signal were then checked against these specified ranges. Furthermore, whether the recorded maxima corresponding to the $F=2 \to F^\prime=3$, $F=2 \to F^\prime=\text{CO} 2,3$, and $F=2 \to F^\prime=\text{CO} 1,3$ transition signals were sequentially increasing, and whether the minima were sequentially decreasing, were verified to identify the desired spectrum.

\begin{figure}[h]
\includegraphics[width=\columnwidth]{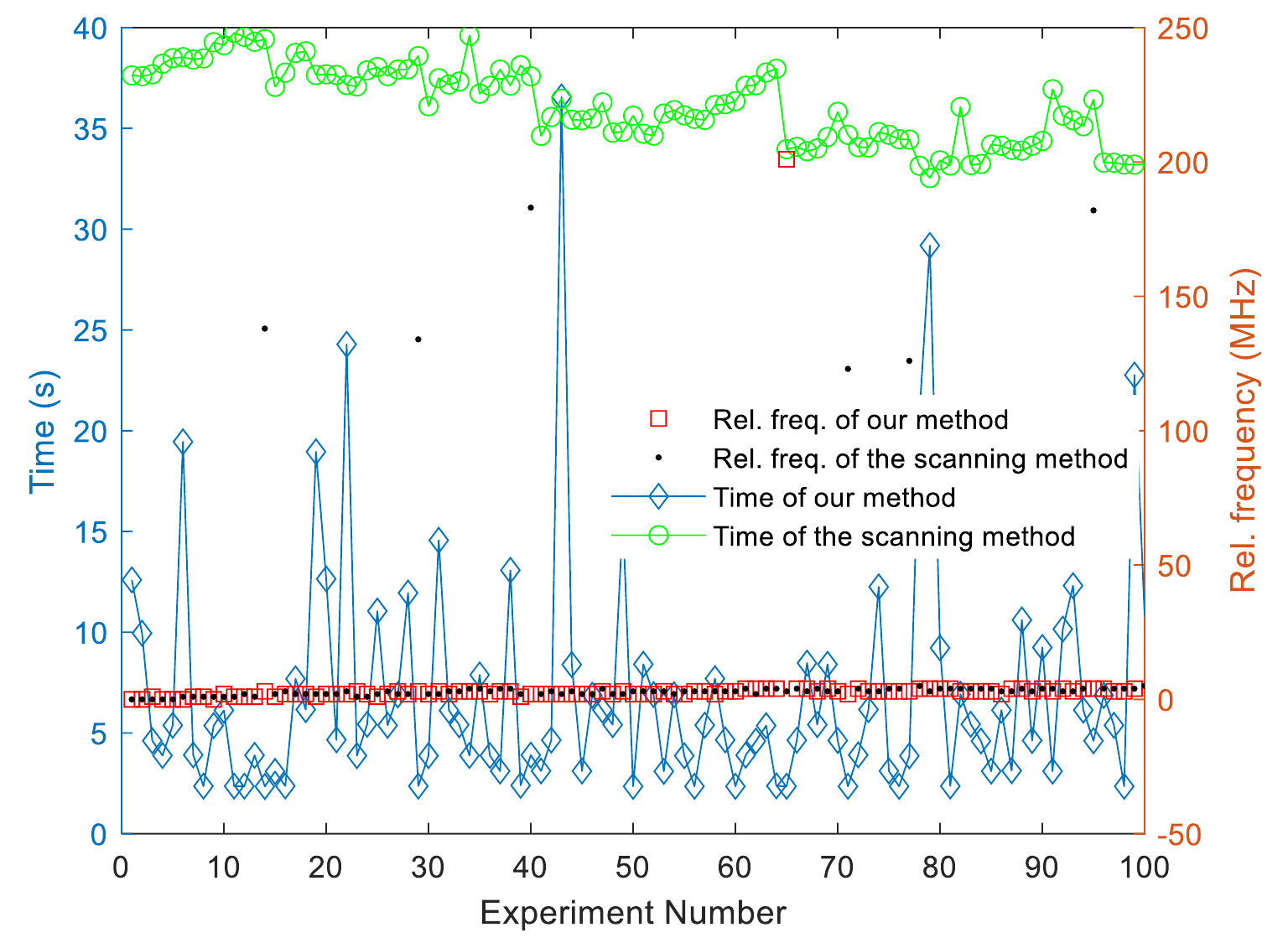}
\caption{\label{fig:fastLocking}(Color online) Comparison of searching time and frequency-locking result between the Bayesian-optimization and gradual-scanning methods. The red squares and black dots indicate the laser-frequency differences between the locked and preset reference for our method and the gradual-scanning method, respectively. The blue line with diamonds and the green line with circles show the desired reference searching time for our method and the gradual-scanning method, respectively.}
\end{figure}

Rapid laser-frequency locking was demonstrated using the Gaussian-process-based Bayesian-optimization and discrete-wavelet-transformation algorithms. The laser-frequency range was scanned by the PZT from 384.2252 to 384.2285~THz, ensuring that no mode hopping occurred within this range. This 3.3 GHz range was discretized into 65 samples with a PZT step of 0.1 V, simulating  the challenging field scenarios, such as severe environmental disturbances or long-term operation, which require a wide-range search to recover the reference signal when the laser frequency drifts far away from the reference point. Two spectrum-searching methods were tested 100~times, with the frequency reference set to 384.227995~THz, corresponding to the $F = 2 \to F' = 3$ transition frequency. The desired spectrum was identified using the wavelet-transformation-based method. Figure~\ref{fig:fastLocking} shows the difference between the laser frequency read from the wavelength meter and the reference frequency set according to the experimental requirement after the laser frequency was locked. Both methods successfully found the desired spectrum within a 3.3-GHz frequency range in all 100~tests. The red squares show the frequency differences obtained using the Gaussian-process-based Bayesian-optimization method, and the black dots show the frequency differences obtained by the gradual-scanning method. There were seven instances in which the locked laser frequency deviated significantly from the reference. These discrepancies can be attributed to the communication rate of the DigiLock~110 because it took 100~ms to enable the proportional-integral-derivative (PID) controller to lock the laser frequency after defining the PID setpoint. During this interval, the MTS signal may have changed, preventing the PID controller from immediately locking on to the zero-crossing point of the MTS signal. 
It can be seen from Fig.~\ref{fig:fastLocking} that the Bayesian-optimization method has superior performance when compared to the gradual-parameter-scanning approach, achieving better results in 99 out of 100 experimental trials. 
This improved performance results from our method's ability to make full use of historical observation 
data within the Gaussian-process-based Bayesian-optimization framework, enabling intelligent prediction of the most promising parameters for reference identification. 
In contrast, the gradual-scanning method relies on brute-force parameter exploration from initial to target values. The experimental results indicate that the average number of PZT parameter-searching iterations was 5.52 for Bayesian optimization, compared to 44.76 for gradual scanning, quantitatively validating the much higher efficiency of our proposed approach. The average search times for the two methods were 7.08 and 36.15~s, respectively, indicating a fivefold improvement in search efficiency compared to the gradual-scanning method. 
To further compare the searching efficiency of our method with that of the gradual-scanning method using the optimized PZT step, a 2.6 GHz mode-hop-free range was discretized into 11 samples with a PZT step of 0.5 V for accelerating the searching speed. This step can also safely capture the three atomic transitions without skipping the desired spectrum. The gradual-scanning method under the optimized condition required an average of only 9.12 iterations (8.7 s). 
Remarkably, the Bayesian optimization method still demonstrated superior efficiency, requiring an average of only 5.02 iterations (3.8 s). This confirms that the Bayesian-optimization method maintains an algorithmic advantage in search efficiency under the conditions of both a fine grid and a coarse grid. While generally being rapid, the Bayesian-optimization algorithm required an extended search time in 6 out of the 100 trials in Fig.~\ref{fig:fastLocking}. These specific instances arise when the algorithm commands a large, instantaneous voltage step, inducing transient mechanical ringing in the PZT actuator. This ringing distorts the collected MTS signal and leads to a false-negative identification.
Because the Bayesian-optimization acquisition function is tuned to favor exploitation, the algorithm temporarily searches other suboptimal regions before eventually returning to explore the true reference, thereby increasing the total search time for these specific trials. 
\begin{figure}[hb]
\includegraphics[width=\columnwidth]{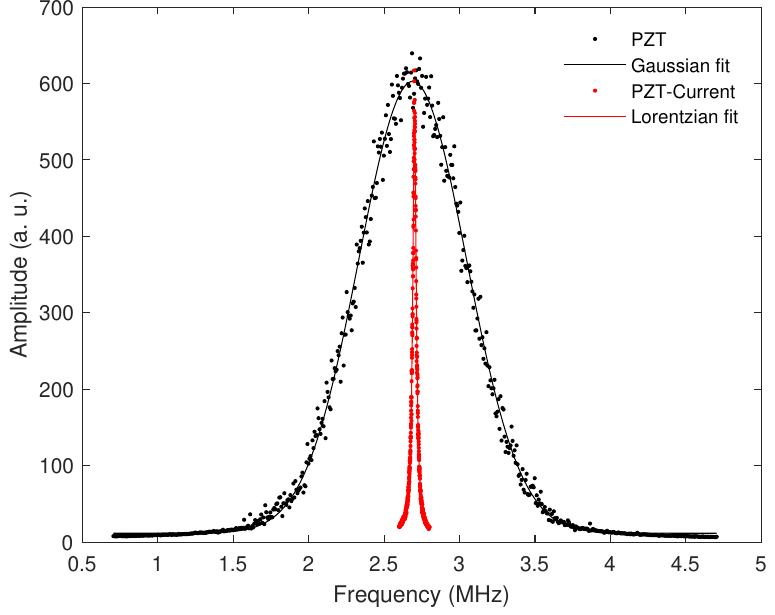}
\caption{\label{fig:linewidth}(Color online) Comparison of the line shapes of the PZT-locking and PZT-current double-loop locking methods. The FWHH of the beatnote signal is narrowed from 840 to 28~kHz when the PZT-current locking method is applied.}
\end{figure}
\begin{figure}[htb]
\includegraphics[width=\columnwidth]{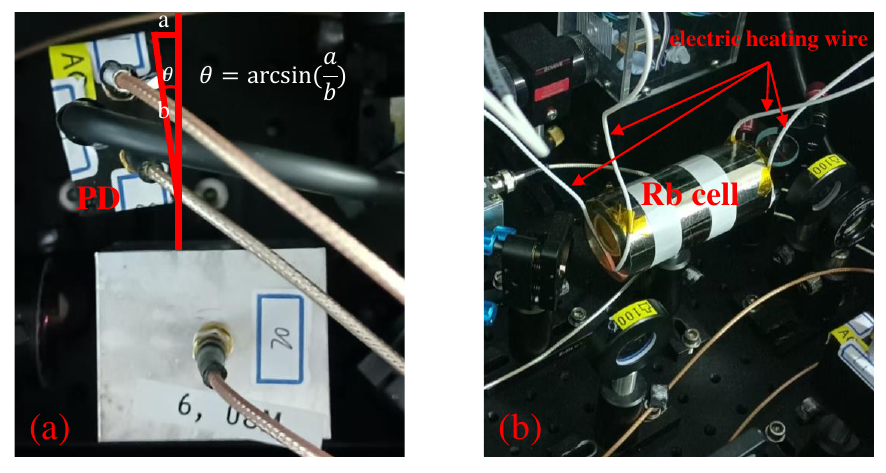}
\caption{\label{fig:robustnessSetup}Experiment setups for alignments of PD (a) and temperature adjustments of Rb cell (b). Different alignment angles of PD are estimated by the formula $\theta=arcsin\left(a/b\right)$ using the measured lengths $a$ and $b$. The Rb cell is heated by the electric heating wires. }
\end{figure}
\begin{figure*}[htb]
\includegraphics[scale=0.40]{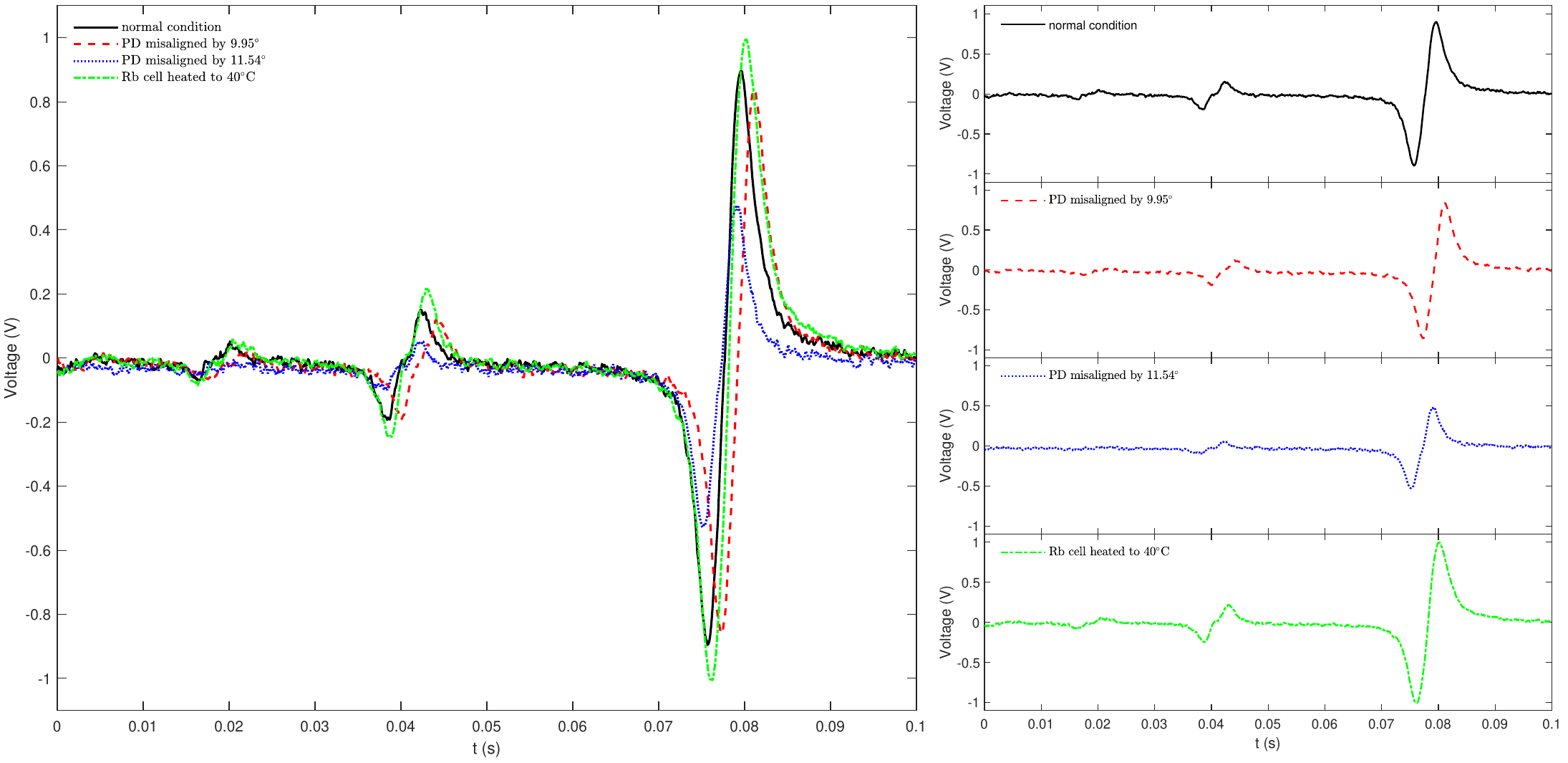}
\caption{\label{fig:robustness}
(Color online) Comparison of MTS signals under different PD misalignments and Rb cell temperatures. The left panel shows the overlaid signals to display their disparities, while the right panel shows each signal individually to distinguish their details. }
\end{figure*}
It is also worth noting that in Fig.~\ref{fig:fastLocking}, the search time for the gradual-scanning method exhibits an overall decreasing trend across the 100 experimental trials, 
because the voltage coordinates corresponding to the atomic transitions drift closely to the start of the search range. The proposed auto-locking scheme consistently and accurately identified the desired spectrum. 
This result demonstrates advantage of the discrete-wavelet identifier, i.e., by evaluating the immutable relative frequency differences and spectral morphology of the transitions rather than relying on absolute voltage lock-points. The laser frequency was locked after the reference was autonomously sought and identified. As the black data points and black fit line in Fig.~\ref{fig:linewidth} show, the full width at half height (FWHH) of the beatnote signal is 840~kHz with only the PZT feedback loop, measured by the heterodyne beatnote method using two identical, independently locked diode lasers. Because this 840 kHz linewidth is largely dominated by the free-running laser frequency noise, it is difficult to meet the experimental demand of tens of kilohertz in the sensitivity-limit measurement~\cite{mcguirk2001low,Biedermann2009}. 
Therefore, the servo loop bandwidth must be expanded into the megahertz regime. To further reduce the linewidth, the bandwidth of the feedback loop was expanded by adding a laser-current feedback loop with a bandwidth of 20~MHz. With this double feedback loop, the high-frequency noise was suppressed, and the FWHH was significantly narrowed from 840 to 28~kHz, as the red data points and red fit line in Fig.~\ref{fig:linewidth} show, implying that the single laser linewidth is below 20~kHz.

The robustness of the auto-locking scheme was tested by changing the laser power. It was first increased from 17.6 to 19.4~mW, before being decreased to 8.8~mW. Correspondingly, the maximum voltage of the desired spectrum first increased by 1.14\% and then decreased by 25.79\%. Under each laser-power condition, 1000~MTS error signals were collected for testing. During the testing process, if the desired spectrum was not identified at the preset scale factor, an attempt was made to identify the reference signal at the preceding or subsequent scale factors in the wavelet decomposition. If the desired spectrum still could not be successfully identified, the zero-crossing points of the spectrum obtained at the current and previous scale factors were combined to determine whether the current signal corresponded to the desired spectrum. The test results indicate that the identification rate of the desired spectrum varies with laser power. Nonetheless, the proposed method achieved a minimum accuracy of 99.8\%, with the highest identification rate reaching 99.9\%, validating the robustness of our approach. There are three possible reasons for false-negative spectrum identification. First, the selected 15 maximum-value points of the detail function either did not contain---or only contained one or two---maximum-value points corresponding to the $F=2 \to F^\prime=\text{CO} 1,3$ transition signal. Even when two maximum points were present, the frequency difference between them exceeded the preset frequency range and they were thus not clustered together. Second, the maximum-value points corresponding to the $F=2 \to F^\prime=\text{CO} 1,3$ and $F=2 \to F^\prime=\text{CO} 2,3$ transition signals were accidentally clustered together due to spectrum noise. Third, the clustering of maximum values of the detail function corresponding to the $F=2 \to F^\prime=\text{CO} 2,3$ transition signal is also affected by spectrum noise, shifting the search range for zero-crossing points of the transition signal. 
To further verify the robustness of the proposed method, the algorithm is evaluated by adjusting the PD alignment and the Rb cell temperature, as shown in Fig.~\ref{fig:robustnessSetup}. First, the PD was deliberately misaligned by angles of $2.83^\circ$, $5.67^\circ$, $9.95^\circ$, and $11.54^\circ$. Under each condition, 1,000 target MTS signals were recorded and evaluated, as shown in Fig.~\ref{fig:robustness}. 
Although the shape and amplitude of the spectrum were changed, the high identification accuracies of $99.7\%,\;99.5\%,\;99.7\%$ for the first three angles were maintained, demonstrating high robustness up to a deviation of $10^\circ$. At the misalignment of $11.54^\circ$, the accuracy dropped to $53.6\%$, because severe clipping of the probe beam on the PD active area limits the light exposure, causing the weakest spectral feature, i.e., $F=2 \to F^\prime=\text{CO} 1,3$ transition, to be buried in the detector noise. Similarly, the robustness can also be verified by changing the spectral shape and amplitude through adjusting the misalignment between the pump and probe beams. 
Second, the Rb vapor cell was heated from $22^{\circ}$C to $40^{\circ}$C. As shown in Fig.~\ref{fig:robustness}, for the 1000 signals collected under the heated Rb cell condition, the identification rate reached $100\%$. This excellent performance is attributed to the higher atomic vapor density caused by the increased temperature, which naturally enhances the SNR of the MTS error signal. To evaluate the false-positive rate of the proposed method, 1000~frames of data for each non-desired spectrum were collected and analyzed, as illustrated in Figs.~\ref{fig:intensityVariations}(d) and~\ref{fig:intensityVariations}(e). These results show that there are no false-positive cases. Overall, these test results demonstrate that our method exhibits a highly successful identification rate with the strong robustness against variations of laser power, Rb cell temperature, and PD alignment.

\section{\label{sec:Conclusion}CONCLUSION AND DISCUSSION}

In this paper, we have proposed and demonstrated a rapid and robust laser-frequency auto-locking method with a narrowed linewidth. A Gaussian-process--Bayesian-optimization algorithm was developed to rapidly search for the reference signal by making use of the observed data to purposefully select the next parameter for finding the desired reference, rather than blindly scanning the parameters step by step. A discrete-wavelet-transformation algorithm was used to locate the transition signals in the collected signals in real time. By analyzing the frequency differences and relative magnitudes of the transition signals, which are determined by the atomic energy-level structure and the relative atomic transition probabilities, the reference signal was robustly identified. After the reference signal was rapidly located and robustly identified, the laser frequency was auto-locked using the PZT-current dual-loop servo. The experimental results showed that our method achieves a fivefold improvement in search efficiency compared to the gradual-scanning method when the laser frequency drifts far away from the reference, along with 99.5\% identification accuracy, even under 50\% laser power fluctuations, $9.95^\circ$ photodiode misalignment, and $18^\circ$C Rb cell temperature elevation, validating its high robustness. Using the heterodyne beatnote method, the laser linewidth was found to be 20~kHz. Although the approach was demonstrated here for MTS lines, its universality enables straightforward extension to other spectral references (e.g., SAS lines). 
In the future, the system will be tested through vibration, shock, and variations in pressure, humidity, and temperature for field applications. Furthermore, we will explore integrating machine-learning-based noise profiling to autonomously diagnose mode hops.

\vspace{6pt}

\begin{acknowledgments}
We acknowledge the financial support provided by the National Key Research and Development Program of China under Grant No.~2025YFF0515200; the Quantum Science and Technology-National Science and Technology Major Project of China under Grant No.~2021ZD0300604; the National Natural Science Foundation of China under Grants 
No.~U25D9005, No.~12104466, and No.~12504571; the Postdoctoral Innovation Research Posts in Hubei Province of China under Grant No.~R20R0004; the Natural Science Foundation of Wuhan under Grant No.~2025040601030117; the Shenzhen Fundamental Research (Key Program) under Grant No.~JCYJ20241202124931042; the Key Science and Technology Project of the Shenzhen Science and Technology Innovation Commission (SSTIC) under Grant No.~KJZD20230923115505011; the cluster special project of Shenzhen Institutes of Advanced Technology, Chinese Academy of Sciences under Grant JQ0101-2025/02.
\end{acknowledgments}


\section*{Disclosures}
The authors declare no conflicts of interest.

\section*{Data Availability Statement}
All relevant data are available from the authors upon request.

\section*{References}
\providecommand{\noopsort}[1]{}\providecommand{\singleletter}[1]{#1}%

\end{document}